\documentclass[a4,12pt]{article}
\usepackage{bm}
\usepackage{amsmath}
\usepackage{epsfig}
\begin{document}
\begin{center}
{\bf \large Two-pion production processes, chiral symmetry and\\ $NN$ interaction in the medium.}\\[2ex]

G. Chanfray$^1$, D. Davesne$^1$, M. Ericson$^{1,2}$ and M. Martini$^1$\\

$^1$ IPN Lyon, IN2P3-CNRS et UCB Lyon I, F-69622 Villeurbanne Cedex\\

$^2$ Theory division, Physics Department, CERN, CH-12111 Geneva 

\begin{abstract}
We study the two-pion propagator in the nuclear 
medium. This quantity appears in the $\pi\pi$ T-matrix and we show that
 it also  enters the QCD scalar susceptibility.
 The  medium effects on this propagator are due to the influence of the
individual nucleon response to a scalar field through their pion clouds. This 
response is appreciably increased by the nuclear environment.  It produces an
 important convergence effect between the scalar and pseudoscalar susceptibilities,
reflecting the reshaping of the scalar strength observed in $2\pi$
production experiments.  While a large modifications of the $\sigma $
propagator follows, due to its coupling to two pion states, we show that the NN
 potential remains instead unaffected.
\end{abstract}
Pacs: 24.85.+p 11.30.Rd  12.40.Yx 13.75.Cs 21.30.-x
\end{center}

 \section{Introduction}
The  two-pion production experiments  on nuclei 
[1-4] have revealed a 
striking accumulation of scalar strength for the $2\pi$ invariant mass near threshold,  
restricted to the isoscalar channel for the two pions.
 Schuck {\it et al.} \cite{SNC88} and Chanfray {\it et al.} \cite{CASN91} have predicted
 such effects on the basis of the influence  of the modification of the 
pion dispersion relation in the  medium  on the scalar
 strength distribution. The pion line
is replaced by a pion branch, a collective mixture of pions and $\Delta$-hole states, 
which lies at lower energies. It leads to a concentration of strength 
near the  $2\pi$ threshold for  the sigma meson which decays in two pions. 
Other interpretations have been 
given in Refs. 
[7-9]. Focusing on the interpretation of 
Refs.\cite{SNC88} and \cite{CASN91}, one of our aims
 is to establish the existence of a link, albeit not a 
straightforward one, between the softening of the scalar strength
 and chiral symmetry restoration. The link goes as follows: the enhancement of the $\pi\pi$ 
 T-matrix near threshold arises from the nuclear modification of the two-pion propagator. 
 This also affect the QCD scalar susceptibility, which will be our main topic.\\
 Beside 2$\pi$ states, the lowest excitations modes of the vacuum which 
govern the scalar susceptibility can also imply 
 a genuine scalar isoscalar meson, the so called ``sigma'' meson.
 These aspects can be incorporated in the linear sigma model, with a $\sigma$ chiral 
partner of the pion strongly coupled
 to 2$\pi$ states. We will discuss also how the propagation of the sigma meson is affected by 
 the modification of the two-pion propagator.

The other point that we will elucidate is the connection with 
traditional aspects of the nuclear binding since the strong in-medium reshaping 
of the scalar strength has {\it a priori} consequences for this problem. 
Our discussion will be based on the distinction, previously emphasized \cite{CEG02},  
between the sigma, chiral partner of the pion, and the scalar meson exchange of 
nuclear physics.

\medskip
Our article is organized as follows. In section 2 
starting from the expression of the $\pi\pi$ T-matrix, 
we introduce the effect of the two-pion propagator.
 
We then focus on 
the zero four momentum case and  we establish the physical interpretation 
of the medium effects, 
linking them to QCD properties. We pursue in section 3 with the discussion 
of the implications for the  nuclear binding. Section 4 is devoted to 
comments concerning the scalar field propagation and to the concluding remarks.

 \section{T-matrix, two-pion propagator and scalar susceptibility in the medium}

In the $2\pi$ production experiments the medium effects are governed by 
the in-medium modifications of the T-matrix for $\pi\pi$ scattering. 
In the following we will study this quantity in a chiral model, the linear sigma one.
In fact our results on the medium effects do not depend on the scalar meson mass 
but essentially on the two-pion propagator that could be obtained directly in the non 
linear sigma model. It is interesting however to keep a sigma meson with a finite mass 
in order to establish link with the binding properties of nuclei.

In this model the coupling of
the $\sigma$ to two-pion states is a simple contact interaction, $\lambda 
f_\pi\,\sigma\pi^2$, while the four pion interaction is $\lambda\pi^4/4$. 
The coupling constant  $\lambda$ is related to observables according to~: 
$\lambda=(m^2_\sigma-\,m^2_\pi)/2 f_\pi^2$. The corresponding $s$-channel 
contribution, $V_s$, to the $\pi\pi$ potential at a 
given invariant  squared mass $s$ is the sum of a contact term and a sigma exchange one. 
The same structure also holds for  the $t$ and $u$ channels. We use an 
approximation suggested by Aouissat \textit{et  al.}~\cite{ASW97}, who keep only the 
$s$-channel term, dropping the $t$ and $u$ channel 
contributions which enter with a smaller weight in the isoscalar channel.
These authors have shown that, within a  symmetry conserving $1/N$ expansion 
(here $N=4$) fulfilling Ward identities, this is a legitimate 
approximation. Within this simplified framework they were able to reproduce 
the $\pi\pi$ phase shifts and scattering length . Then the
scalar-isoscalar potential reads~:         
\begin{equation}
V=6\,\lambda\,+\, 24\,(\lambda\,f_\pi)^2\,{1\over s\,-\,m^2_\sigma}=
6\,\lambda\,{E^2\,-\,m^2_\pi\over E^2\,-\,m^2_\sigma},
\label {V}
\end{equation}
being $E=\sqrt{s}$ the total  energy of the pion pair in the CM frame.\\
The Lippmann-Schwinger equation with such a separable potential
 gives for the unitarized scalar-isoscalar T-matrix ~:
\begin{equation}
T(E)=
{6 \lambda (E^2-m^2_\pi)\over E^2-m^2_\sigma\,- \,3 \lambda\, 
(E^2-m^2_\pi)\,G(E)}\,,
\label{TMAT}
\end{equation}
which in the limit of large sigma mass reduces to (assuming zero three momentum for the 
pion pair)~:
\begin{equation}
T(E)=-
{3\,(E^2-m^2_\pi)/ \,f_\pi^2 \over 1\,+\, 
3\,(E^2-m^2_\pi)/(2\,f_\pi^2)\,G(E)}\,.
\label{TMATUN}
\end{equation}
Here $G(E)$ is the two-pion propagator for zero three momentum 
of the pion pair, linked to the single pion 
propagator  $D_\pi$ by~:  
\begin{equation}
G(E)=\int {d{\bf q}\over (2\pi)^3}\,
\int{i \,dq_0\over 2\pi}\, D_\pi({\bf q}, q_0)\, D_\pi(-{\bf q},\, E-q_0).
\label{TWOPI}
\end{equation}

The  expression of $T(E)$ given in Eq. (\ref{TMAT}) holds  either in the 
vacuum or in the medium. In the last case the pion propagators which enter $G$ are 
dressed by  particle-hole bubbles, where  the particle is either a nucleon or a Delta.  
 This is
 responsible   for the modification of the T-matrix in the medium and, in the
  interpretation of Refs. \cite{SNC88,CASN91,ROV02}, explains the features of the data.

We now come to the link with the chiral symmetry. The two-pion propagator is the
 correlator of a scalar quantity, the squared pion field. On the other hand 
the order parameter for 
the spontaneous breaking of the chiral symmetry is another scalar, the quark condensate, 
that we denote $\langle\bar{q}q(\rho)\rangle$ in the
nuclear medium, whereas $\Delta\langle\bar{q}q(\rho)\rangle$ is the variation 
with respect to the vacuum. The QCD scalar susceptibility, defined as the derivative of the quark condensate  with respect 
to the quark mass, is the correlator of the fluctuation of the quark scalar density. 
We define the nuclear value $\chi^A_S$ as 
the difference with the vacuum quantity~:
\begin{equation}
\chi^A_S(\rho)=\frac{d}{d m_q} \Delta\langle\bar{q}q(\rho)\rangle.
\label{ACHIS}
\end{equation}
One can expect a link between the two scalar correlators: the QCD susceptibility and 
the two pion propagator. Indeed,
a major contributor to chiral symmetry restoration is the nuclear pion cloud. 
In the non-linear sigma model this is the only agent for restoration,
each pion of the cloud contributing to the restoration by an amount proportional to
$\Sigma_{\pi}=m_{\pi}/2$, sigma commutator of the pion, the pionic 
participation to the restoration is expressed
 in terms of the pion density 
in the cloud, $m_\pi\langle\Phi^2\rangle$, according to~\cite{ChEr93}~:
\begin{equation}
\Delta\langle\bar{q}q(\rho)\rangle = - \langle\bar{q}q\rangle_{vac}
\frac {\langle\Phi^2\rangle}{2 f_{\pi}^2}.
\label{DELTAQBARQ}
\end{equation}
The quantity $\langle\Phi^2\rangle$ is related to the pion 
propagator $D_{\pi}(q)$ by~ :
\begin{equation}
\langle\Phi^2\rangle=3\int \frac{i~d^4q}{(2\pi)^4}~[D_{\pi}(q)-D_{0\pi}(q)]=
3\int \frac{i~d^4q}{(2\pi)^4}~\left[\frac{1}{q^2-m_{\pi}^2-S_{\pi}(q)}-
\frac{1}{q^2-m_{\pi}^2} \right].
\label{PHI2}
\end{equation}
In the expression above the vacuum value of the pion propagator $D_{0\pi}(q)$ is 
subtracted out in order to retain only medium effects and 
 $D_{\pi}(q)$ is related to the pion self-energy $S_{\pi}(q)$, 
which includes an $s$-wave, $S_s$, and a $p$-wave part, $S_p$. 
The second piece arises from
 the $p$-wave excitations of particle-hole, yielding a three-momentum dependent coupling of the pion.
 It does not depend explicitly 
on the pion mass, while the $s$-wave one does.

For the evaluation of the susceptibility the derivative with 
respect to the quark 
mass is replaced as usual by the one with  respect to the pion mass squared. 
 We ignore the derivative of the 
 $s$-wave potential which leads to small corrective terms. 
With this approximation we 
obtain a simple expression for the susceptibility~:

\begin{eqnarray}
\chi^A_S(\rho)=\frac{\langle\bar{q}q\rangle^2_{vac}}{f_{\pi}^4}\frac{d}
{d m_{\pi}^2}\langle\Phi^2\rangle
=3\frac{\langle\bar{q}q\rangle^2_{vac}}{f_{\pi}^4}
\int \frac{i~d^4q}{(2\pi)^4}~[D_{\pi}^2(q)-D_{0\pi}^2(q)]
=3\frac{\langle\bar{q}q\rangle^2_{vac}}{f_{\pi}^4}  \Delta G(0)
\label{ACHIDELTAG}
\end{eqnarray}
which is proportional to $ \Delta G(0)$, the in-medium modification, at E=0, of the 
two-pion propagator, the quantity which governs the modification of the $\pi\pi$ T-matrix.

In order to illustrate the significance of this 
contribution we first consider a  single insertion of a nucleon-hole bubble 
 (namely $\Pi_N^0$) into one of the 
two-pion lines (see Fig. 1). The corresponding medium correction reads~:

\begin{equation}
\chi^A_S(\rho)=3\frac{2\langle\bar{q}q\rangle^2_{vac}}{f_{\pi}^4}\,
\int{i \,d^4q\over (2\pi)^4}\,
D_{0\pi}(-q)\,D_{0\pi}^2(q)\,{\bf q}^2\,\Pi_N^0(q).
\end{equation}

The physical interpretation   
follows from Fig 1~: this medium correction
 introduces  the effect  of the individual nucleonic susceptibility  from 
their pion clouds. Note that the Pauli blocking effect is 
implicitly contained 
through the quantity $\Pi_N^0$. Ignoring it, the contribution to 
the scalar susceptibility is $\rho_s\chi_S^N(\pi)$ where $\rho_s$ is the 
nucleon scalar density and we 
have denoted $\chi_S^N(\pi)$
 the free nucleon susceptibility from its pion cloud.   
 This quantity was discussed by  Chanfray \textit{et al.} \cite{CEG03} who
showed  that it dominates the nucleon response and who 
evaluated it in the static 
approximation: $\chi_S^N~=- 4. 10^{-2}$ MeV$^{-1}$.

\begin{figure}
\begin{center}
\epsfig{file=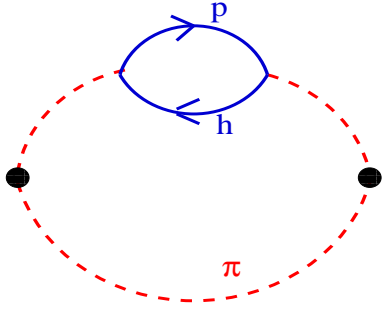,width=5.0cm,height=4.0cm,angle=0}
\end{center}
  \caption{ Influence of the nucleonic pion cloud on the two-pion propagator. For the susceptibility the quark fluctuation is 
attached at each point.}
\label{}      
\end{figure}
In the description above we have considered in the two-pion propagator the 
dressing of a single pion line by
 only one bubble. If instead we introduce in the two-pion propagator the full 
RPA pion propagator in the ring approximation, the 
contribution to the nuclear susceptibility 
can be written as  $\rho_s\tilde{\chi }_S^N$, 
with an in-medium modified, density dependent
 nucleonic scalar susceptibility $\tilde{\chi }_S^N$. 
The relation between $\Delta G$ and $\tilde{\chi }_S^N$ is~:
\begin{equation}
3\Delta G(0)=\frac{\rho_s \tilde{\chi }_S^N f_{\pi}^4}{\langle\bar{q}q\rangle^2_{vac}}.
\end{equation}
At $\rho_0$ it happens that $3\Delta G(0)\simeq\tilde{\chi }_S^N$, if $\tilde{\chi }_S^N$ 
is expressed 
in MeV$^{-1}$. 
For its evaluation we proceed as follows.
First, the complete bare polarization propagator $\Pi^0$ is the sum of 
the nucleon-hole polarization propagator $\Pi^0_N(q)$ and of the 
delta-hole one, namely
\begin{eqnarray}
  \Pi^{0}_{\Delta}({\bf q},\omega)& =&
\frac{4}{9} \left(\frac{g_A}{f_{\pi}}\right)^2
\left(\frac{g_{\pi N\Delta}}{g_{\pi NN}}\right)^2
F^2({\bf q},\omega)
   \int\frac{\mbox{d}{\bf k}}{(2\pi)^3}\theta(k_F-k)\nonumber \\
&&\left(\frac{1}  
{\omega+\epsilon_{{\bf k}}-\epsilon_{\Delta,{\bf k}+{\bf q}}
+i\Gamma_{\Delta}({\bf k}+{\bf q},\omega)}-
\frac{1}{\omega+\epsilon_{\Delta,{\bf k}-{\bf q}}-\epsilon_{{\bf k}}}
    \right),
\end{eqnarray}
where $g_{\pi NN}$ ($g_{\pi N\Delta}$) is the $\pi NN$ 
($\pi N\Delta$) coupling constant, $F(\bf{q},\omega)$ is the form factor at the $\pi NN$
 or $\pi N\Delta$ vertex  
and $\Gamma_{\Delta}$ is the 
delta width (taken following Ref.\cite{Ose87}). Moreover we have defined:
$\epsilon_{{\bf k}}={\bf k}^2/2 M_N$ and $\epsilon_{\Delta,{\bf k}}=
M_{\Delta}-M_N\,+\,({\bf k}^2/2 M_{\Delta})$.\\
The ($p$-wave) pion self-energy  is linked to the fully dressed polarization propagator, 
$\tilde{\Pi}$, solution of the RPA equations in the ring 
approximation, as follows~:
\begin{eqnarray}
S_\pi({\bf q},\omega)
={\bf q}^2 \tilde{\Pi}({\bf q},\omega)={\bf q}^2
\frac{\Pi^{0}_{N}({\bf q},\omega)+\Pi^{0}_{\Delta}({\bf q},\omega)-
(g'_{NN}+g'_{\Delta\Delta}-2g'_{N\Delta})
\Pi^{0}_{N}({\bf q},\omega)\Pi^{0}_{\Delta}({\bf q},\omega)}
{[1-g'_{NN}\Pi^{0}_{N}({\bf q},\omega)]
[1-g'_{\Delta\Delta}\Pi^{0}_{\Delta}({\bf q},\omega)]
-g'^2_{N\Delta}
\Pi^{0}_{N}({\bf q},\omega)\Pi^{0}_{\Delta}({\bf q},\omega)}
\end{eqnarray}
where $g'_{NN}$, $g'_{\Delta\Delta}$, $g'_{N\Delta}$ are the Landau-Migdal 
parameters for the $NN$ channel, for the $\Delta\Delta$ one and for the mixing 
of $NN$ and $\Delta N$ excitations, respectively.\\
Inserting this expression of the pion self-energy into the two-pion propagators of Eq. 
(\ref{ACHIDELTAG}) we obtain
 the  nuclear susceptibility, hence the effective nucleonic one. 
It is interesting to compare the 
effective and the free nucleon susceptibilities. For the last quantity we want to  
avoid the static approximation of Chanfray {\it et al.} \cite{CEG03}. 
We can use the same expression (\ref{ACHIDELTAG}) in the dilute limit, $\rho\to0$ so as to eliminate
 the influence of the medium on the susceptibility. We have also
 introduced another method starting from the general expression of the pion density in terms of the spin-isospin 
longitudinal response function, $R_L$, 
as given in Ref. \cite{Chanfray:1999nn}, 
\begin{equation}
\label{PHI2RL}
\langle\Phi^2\rangle=\frac{3 \rho}{A}
\int {d{\bf q}\over (2\pi)^3}\,
\int_{0}^{\infty}{d \omega} 
\left(\frac{1}{2 \omega_q^2 (\omega+\omega_q)^2}+
\frac{1}{2 \omega_q^3 (\omega+\omega_q)}\right)R_L({\bf q},\omega)
\end{equation}
with $\omega_q=\sqrt{{\bf q}^2+m_{\pi}^2}$. The response $R_L$ is~: 
\begin{equation} 
R_L({\bf q},\omega)=-\frac{V}{\pi} Im \Pi_L({\bf q},\omega)
=-\frac{V}{\pi} Im \left({\bf q}^2\tilde{\Pi}({\bf q},\omega)
\frac{\omega^2-\omega_q^2}{\omega^2-\omega_q^2-
{\bf q}^2\tilde{\Pi}({\bf q},\omega)}\right).
\end{equation}
Inserting this expression (\ref{PHI2RL}) of $\langle\Phi^2\rangle$ into the pion condensate and taking the derivative
 with respect to the pion mass squared 
 gives the nuclear susceptibility. 
The expression (\ref{PHI2RL}) holds  in the 
dense case as well as in the dilute limit, \textit{i.e.}, for an assembly of
independent nucleons. In the latter case the expression of the response simplifies to~: 
\begin{equation} 
R_L({\bf q},\omega)=\left(\frac{g_A}{f_{\pi}}\right)^2{\bf q}^2
F^2({\bf q})\left[\delta(\omega-\epsilon_{{\bf q}})-\frac{1}{\pi}
\frac{4}{9}\left(\frac{g_{\pi N\Delta}}{g_{\pi NN}}\right)^2
\textrm{Im} \frac{1}{\omega-\epsilon_{\Delta,{\bf q}} +i\Gamma_{\Delta}} \right]
\end{equation}

and the expression of the free nucleon susceptibility becomes~: 
\begin{eqnarray}
\label{RHO_TO_ZERO}
\chi^N_S&=&-\frac{3}{16 \pi^2}
\frac{\langle\bar{q}q\rangle^2_{vac}}{f_{\pi}^4}
\left(\frac{g_A}{f_{\pi}}\right)^2
\int_0^{\infty} dq~ q^4 F^2(q)
\left\{\frac{3 \epsilon_q^2+9\epsilon_q\omega_q +8\omega_q^2}
{2\omega_q^5(\epsilon_q+\omega_q)^3}
\right.\nonumber\\
&+&\left.
\frac{4}{9}\left(\frac{g_{\pi N\Delta}}{g_{\pi NN}}\right)^2 
\int_0^{\infty} d\omega \left[
\frac{3 \omega^2+9\omega \omega_q +8\omega_q^2}
{2\omega_q^5(\omega+\omega_q)^3}\right]
\left(-\frac{1}{\pi} \textrm{Im} 
\frac{1}{\omega-\epsilon_{\Delta,q}+i\Gamma_{\Delta}} \right)
\right\}.
\end{eqnarray}

We have checked numerically that the first method converges to this free nucleon result
 in the dilute limit.
 Table \ref{tab} shows the results obtained. The free nucleon susceptibility
 calculated from Eq.(\ref{RHO_TO_ZERO}) is displayed in the column $\rho=0$, while the
 other columns display the effective ones
 at normal nuclear
 matter density.  The free value depends on the parameter $\Lambda$ entering 
in the monopole form factor $F({\bf q})=\Lambda^2/(\Lambda^2+{\bf q}^2)$. 
The effective one depends in addition on the $\pi N \Delta$ coupling constant and
 on the Landau-Migdal parameters. We have explored the dependence on these parameters.
  In all cases the magnitude of the susceptibility is appreciably increased as compared to 
the free one \textit{ i.e.} the Pauli blocking is overcompensated by  the effect of the increase of
 the pion propagator in the medium. The enhancement of the effective susceptibility
 is more pronounced with a harder 
form  factor or with smaller Landau-Migdal parameters.

\begin{figure}
\begin{center}
\includegraphics[clip,height=7.0cm]{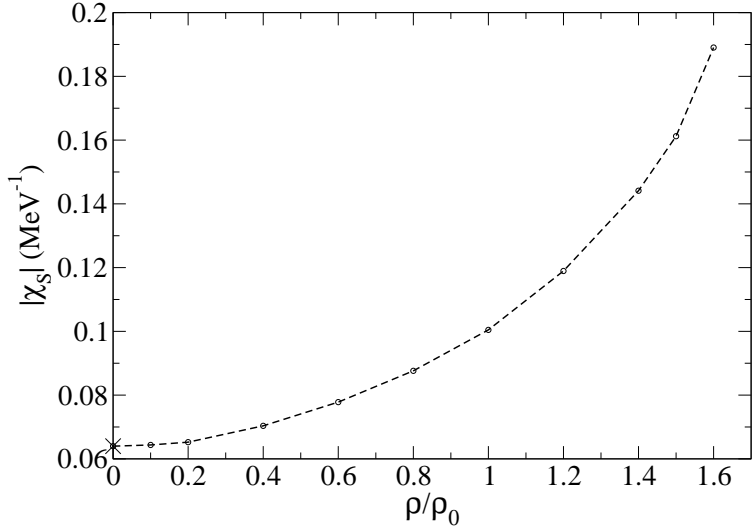}
\caption{\label{fig:chi}Density evolution of the scalar susceptibility 
with the following choice of the parameters: $\Lambda=900$ MeV, 
$\left(\frac{g_{\pi N\Delta}}{g_{\pi NN}}\right)^2=3.8$ and 
$g'_{\Delta}=0.4$. The point at $\rho=0$ is calculated from Eq.(\ref{RHO_TO_ZERO}).}
\end{center}
\end{figure}

Fig. \ref{fig:chi} displays the evolution  with density of the effective nucleon
susceptibility, $\tilde{\chi }_S^N(\pi)$, for a given choice of parameters. As expected it 
converges to the value of 
Eq.(\ref{RHO_TO_ZERO}) in the $\rho \to 0$ 
limit.  Beyond the normal density the increase with density of the effective susceptibility 
becomes rapid.
 
\begin{table}[t]
\begin{center}
\begin{tabular}{|c|ccc|ccc|}
\hline
$\Lambda$ MeV&$\rho=0$&$g'_{\Delta}=0.4$&$g'_{\Delta}=0.5$
&$\rho=0$&$g'_{\Delta}=0.4$&$g'_{\Delta}=0.5$\\
\hline
5 $m_{\pi}$&-5.5&-7.3&-6.5&-5.0&-5.9&-5.4\\
800&-6.0&-8.8&-7.7&-5.4&-7.0&-6.4\\
900&-6.4&-10.0&-8.7&-5.8&-7.8&-7.0\\
1000&-6.7&-11.3&-9.6&-6.1&-8.6&-7.3\\
\hline
&&$\left(\frac{g_{\pi N\Delta}}{g_{\pi NN}}\right)^2=3.8$ 
&&&
$\left(\frac{g_{\pi N\Delta}}{g_{\pi NN}}\right)^2=72/25$ &\\
\hline
\end{tabular}
\caption{\label{tab}The effective nucleonic scalar susceptibility in unit 
of $10^{-2}$ MeV$^{-1}$ at normal nuclear matter density 
compared to the free nucleon one (zero density column)  
for different values of $\Lambda$ and for two values of 
$\left(g_{\pi N\Delta}/g_{\pi NN}\right)^2$ (corresponding respectively to the current nuclear phenomenology 
and to the constituent quark model) 
and $g'_{N\Delta}=g'_{\Delta\Delta}\equiv g'_{\Delta}$. 
The parameter $g'_{NN}$ is kept at the fixed value 
$g'_{NN}=0.7$. } 
\end{center}  
\end{table}

The modification of the scalar susceptibility is linked to chiral symmetry restoration.
 In the phase of spontaneously broken symmetry the susceptibility is 
 split in two, the ``parallel'' one, which is the scalar, and 
the ``perpendicular'', which is the pseudoscalar one \cite{CE03}. The second 
one is infinite in the absence of explicit breaking, \textit{i.e.}, in the limit
$m_{\pi}=0$. The scalar susceptibility instead is driven by hard modes
 (sigma, two-pion \textit{etc.}) and it is much smaller in  magnitude. As the two susceptibilities
 merge in the restored phase
a convergence between them is a sign of a partial chiral
 symmetry restoration .

With increasing density the pseudoscalar susceptibility decreases (in magnitude)
 since, as was established in Ref.\cite{CE03}, it follows the
quark condensate with~:
\begin{eqnarray}
\chi _{PS}(\rho)^À=\frac{\left\langle \overline {q}q (\rho)\right\rangle}{m_{q}^À} 
~~{\rm and } ~~\chi _{PS_{vac}}^À=\frac{\left\langle \overline {q}q \right\rangle_{vac}}
{m_{q}^À}. \label{QCOND}
\end{eqnarray}
At $\rho_{0}$ it has decreased by $\simeq 35\%$.
On the other hand we have seen that the presence of the nucleons which respond through 
their pion clouds contributes to the increase of the scalar susceptibility by (in unit of
 the vacuum value of the pseudoscalar susceptibility)~: 
\begin{equation} 
{\chi _{S}^A\over\chi _{PS_{vac}}}~=~{\rho_S\tilde\chi_S^N~m_q
\over\left\langle \overline {q}q \right\rangle_{vac}}.
\end{equation}
At $\rho_0$ it turns out that numerically this ratio is just  $\tilde\chi_S^N$, with  $\tilde\chi_S^N$
expressed in
 MeV$^{-1}$, which represents a non-negligible $\simeq 8 ~{\rm to} ~10\% $ convergence effect.
 We remind in this respect 
that there exists another and larger factor of convergence which was
studied in Ref. \cite{CEG02}. The conversion of the quark fluctuation density into nucleonic
ones, \textit{i.e.} the effect of the  low lying nuclear excitations on the QCD
scalar susceptibility, produces an additional enhancement of the magnitude of the
 susceptibility, which at $\rho_0$ can be expressed in our unit as~:
\begin{equation}
\chi^A_{S}(\rho_0)/ \chi_{PS_{vac}}~ \simeq~
~9~\rho_0\Sigma_N^2/(f_\pi^2m_\pi^2K),
\end{equation}
where $K \simeq 250~{\rm MeV}$ (close to the free Fermi gas value) 
is the incompressibility factor of nuclear matter
and $\Sigma_{N}¥$ the nucleon sigma commutator. It neglects the relativistic effects at $\rho_0$, in such a way that
 the nuclear responses to a scalar or vector probes can be taken as identical. Moreover 
this result assumes that the nucleon sigma commutator 
is not renormalized in the medium. The
 value of this ratio is $\simeq.56$.
 A more elaborate evaluation by Chanfray \textit {et al.} \cite{AcHu}, which takes into account in particular 
 medium effects in the conversion coefficient, gives for the ratio $\chi_S(\rho_0)/ \chi_{PS_{vac}}$ 
a value $\simeq$ .35.
 Adding the two sources of modification and neglecting the vacuum value of $\chi_S$ 
which is small and depends on the uncertain sigma mass (as a magnitude order 
$\chi_{S_{vac}}~/ \chi_{PS_{vac}}~\simeq .05)$, the scalar
susceptibility at $\rho_{0}$ is~:
\begin{equation}
\chi_S(\rho_0)/ \chi_{PS_{vac}}~\simeq .45,
\end{equation}
 while the pseudoscalar one in the same units is $\simeq .65$. 
The two susceptibilities which are drastically
 different in the vacuum become nearly equal in ordinary nuclear matter, a remarkable convergence
 effect. The reason for the large increase of the scalar susceptibility as compared to
 its vacuum value is the spectrum of scalar excitations in the nuclear medium.
 It encompasses 
 nuclear states and also two-quasi-pion states which extend at lower energies than the bare 
 two-pion states.

 With increasing density the contribution of low-lying
nuclear excitations to the scalar susceptibility does not increase 
further and even decreases, as the nuclear
scalar-isoscalar response becomes collective with a repulsive residual force
\cite{CE04}.On the other hand the 
one from the pionic excitations of the nucleons continues to increase, 
due to the rapid increase 
 of the effective nucleon susceptibility. The overall effect leads a smooth behavior 
(up to $\rho\sim 1.6 \rho_0$). For instance, with the evaluation of Ref. \cite{AcHu} for the nuclear part, 
 the overall susceptibility at $\rho=1.6 \rho_0$ becomes: $\chi_S(1.6 \rho_0)/ \chi_{PS_{vac}}~\simeq .47$, 
quite close to its value at $\rho_0$.
 \section{Sigma propagator and nuclear physics implications}

In nuclear physics the attraction is attributed in part to the exchange
of a scalar field
 between nucleons. The previous description
 of the T-matrix in the linear sigma model incorporates
 a scalar sigma field, chiral partner of the pion.
 It is then  interesting to rewrite the T-matrix 
in the following form which displays the propagator of  the sigma field, $D_\sigma$,  
with the inclusion of its coupling to  $2\pi$ states~:
\begin{eqnarray}
T(E)&=&
{6\, \lambda (E^2-m^2_\pi)\over 1-3\lambda G(E)}\,D_\sigma(E)\nonumber\\
D_\sigma(E)&=&\bigg( E^2-m^2_\sigma- {6 \lambda^2 f^2_\pi G(E)\over 
1-3\lambda G(E)}\bigg)^{-1}.\label{TMATBIS}
\end{eqnarray}

 The physical  interpretation of this 
expression is clear~: the $\sigma$ propagator incorporates its coupling to  
$2\pi$ states dressed by all  rescatterings processes which are driven exclusively by 
the  $4\pi$ contact interaction. At $E=0$ we have~:
\begin{eqnarray}
\label{Dsig0}
- D_\sigma(0) &=& {1\over m^2_\sigma\,+ \,3 \lambda (m^2_\sigma - m^2_\pi) 
{G\over 1-3\lambda G}}=
{1-3\lambda G\over \,m^{2}¥_{\sigma }¥-\,3 \lambda m^2_\pi G}\nonumber\\
&\simeq& \frac{1}{m^2_{\sigma }} \,-\,{3\lambda G\over m^2_\sigma}~.
\end{eqnarray}
The medium correction to $D_{\sigma}¥$ from the coupling of the $\sigma$ to $2\pi$
states is~: 
\begin{eqnarray}
    \Delta D_{\sigma}¥(0)~=~{3\Delta G(0)~ \over 2f_{\pi}^{2}},
    \end{eqnarray}
which represents at $\rho_{0}$ a correction of $\simeq 4\cdot10^{-6}~MeV^{-2}$ . 
For comparison, the bare vacuum value with a sigma mass of 700 MeV 
is $\simeq~2\cdot10^{-6}~ MeV^{-2}$, smaller than the medium contribution. 
\\

Notice that without the $\pi\pi$ rescattering correction the numerical value of $\Delta G$ 
at $\rho_0$ is such that the denominator in the expression of $D_\sigma$ (Eq. (\ref{TMATBIS})) 
would be negative.
The existence of a singularity in the $\sigma $ propagator implies an instability with 
respect to a $2\pi$ isoscalar soft mode, which was discussed by Aouissat \textit{et al}.
\cite{ARCSW95}. These authors have argued that the pion rescattering  effect
(due to  the $4\pi$  contact term)    in the sub-threshold region 
could eliminate the instability, as is the case in our expression (\ref{TMATBIS})
at $E=0$.

We have seen above that the large polarization of the nucleon 
through the pion cloud has a large effect on the $\sigma $ propagation. The following question
naturally  arises~: is the large medium modification of the $\sigma$ propagator reflected
in the $NN$ interaction~?  At first sight it is natural to believe that the scalar 
$NN$ potential is  affected in the same way as the $\sigma $ propagator, which would lead
to strong many-body forces. The answer to the question is 
closely related  to the problem of the  identity between the scalar meson responsible for 
the nuclear binding and the sigma, chiral partner of the pion.
The pure identity between the scalar meson which contributes to the nuclear binding and the
chiral partner of the pion is excluded by chiral  constraints, 
emphasized by Birse \cite{B96}.  It would lead to the presence of a term of order 
$m_\pi$ in the $NN$  interaction, which is  not allowed. Nevertheless it is possible
to describe the $NN$ 
attraction in the linear sigma model as showed by  Chanfray {\it et al.} \cite{CEG02}.
By going from Cartesian to polar coordinates, these authors
 introduced a new scalar field called $S=f_\pi + s$. 
This field is associated with the radius of the chiral circle and it is a chiral 
invariant while  the $\sigma$ field is not. They suggested that the scalar 
meson of nuclear  physics  should be identified with this new scalar field.
The justification will be given  later. More precisely the nuclear 
attraction arises  from the mean (negative) value $\bar s$, and the effective nucleon mass is~:
$ M_N^*= M_N\,+\,g_S \,\bar s$. Actually this new formulation  transforms the original 
linear realization of chiral 
symmetry into a non linear one. Consequently chiral constraints are automatically 
respected, as those mentioned by Birse. Moreover, the coupling constant of the $s$ field 
to the nucleon,  $g_S=M_N/f_\pi\simeq 10$,  is not  incompatible with 
the phenomenology of quantum hadrodynamics \cite{SW86}.

The  passage to polar coordinates  cannot affect the physics. For instance, 
the T-matrix for on-shell pions  must be  independent of the representation. It is 
therefore  interesting to rewrite it in a form which displays the
propagator of the $s$ field, the relevant quantity for nuclear physics. 
For this purpose  we now express  the Lagrangian of the linear sigma model in terms of 
the polar  coordinates, {\it i.e.}, the $s$ field and the new pion field $\phi$ which is 
directly related to the chiral angle, being~:
\begin{equation}
\sigma=(f_\pi + s) \cos \left(\frac{\phi}{f_\pi}\right),
~~~~~~\vec{\pi}=(f_\pi + s)\hat{\phi} \sin \left(\frac{\phi}{f_\pi}\right).
\end{equation}

This Lagrangian has been given in \cite{CEG02} and we 
restrict ourselves to pieces relevant for our purpose.  We first note 
that the bare masses for the $s$ and $\sigma$ fields are the same. For 
the   $4\pi$ contact term  we recover  the standard  non-linear sigma model 
result with contain  derivative terms. In addition we get a $s\pi\pi$ coupling piece 
of the  derivative type~:
\begin{equation}
{\cal L}={s\over 
f_\pi}\left(\partial_\mu\vec\phi\cdot\partial^\mu\vec\phi\,-\,{1\over 
2}\,m^2_\pi\,
\phi^2\right).
\end{equation}
As an illustration, summing contact and s exchange pieces the Born amplitude writes:
\begin{equation}
V(E)={6\over f^2_\pi}\left(-(E^2\,-\,m^2_\pi)\,+\,{(E^2\,-\,m^2_\pi)^2
\over E^2\,-\,m^2_\sigma}\right)\label{VNEW}
\end{equation}
which reproduces the previous result of Eq. (\ref{V}), as expected~; only the decomposition 
has changed. The same holds for the T-matrix. The interest lies in the identification
 of the propagator $D_s$ of the 
chiral invariant scalar field $s$. We rewrite the T-matrix in a form which 
displays the coupling, $E^2 - m^2_\pi$, of the $s$ field to two pions 
and the $\pi\pi$ rescattering through the new contact interaction (first term 
of Eq. (\ref{VNEW})). This new  decomposition  writes~: 
\begin{eqnarray}
T(E)&=&
{6\, \lambda (E^2-m^2_\pi)\over 1+{3 \over 2 
f^2_\pi}\,(E^2-m^2_\pi)\,G(E)}\,D_s(E)\nonumber\\
D_s(E)&=& \bigg( E^2-m^2_\sigma- {3\over 2 
f^2_\pi}\,(E^2-m^2_\pi)^2\,{G(E)\over 
1+ {3\over 2} {E^2-m^2_\pi\over f^2_\pi} G(E)}\bigg)^{-1}.\label{TMATNEW}
\end{eqnarray}
We consider the zero-energy case, 
$E=0$, and compare  the propagators of the $s$ and $\sigma$ fields. They differ 
in an  essential way with  respect to their coupling to two pions. For the $s$, the 
coupling  vanishes  in the  chiral limit, hence it is small and it can be ignored, while 
it is  large  for the $\sigma$. This difference is due to the chiral invariant character 
of the $s$ field.
As the $NN$ scalar potential is linked to the exchange  of the  $s$ mode, 
it is not modified by the medium effects discussed previously,
which affect the $\sigma$ propagation through its coupling to  $2\pi$ states. 

We now come to the
justification of the identification of the nuclear attraction with the $s$ field exchange.
The physics cannot depend on field transformation from Cartesian to polar coordinates. Hence the
same conclusion about  the stability of the $NN$ potential  should be 
reached also in the original linear formulation. In this case the nucleons exchange a $\sigma$ 
with its $\pi\pi$ dressing but the consistency of the model also implies other exchanges 
with resulting delicate compensations \cite{B94}. Their origin is the well-known pair
 suppression, in the case of pseudo-scalar coupling, by
$\sigma$ exchange for the  $\pi N$ amplitude. As depicted in Fig. 3b,  this translates into the 
elimination 
of the sigma dressing in the $NN$ interaction. We have explicitly checked  that this 
cancellation holds to all orders in the dressing of the sigma.   
The net result amounts to the exchange of the $s$ mode and hence to the
identification of Chanfray {\it et al}. \cite{CEG02}. Their
formulation provides a very economical way to incorporate all 
the cancellations inherent to the linear realization, and hence the requirements of 
chiral symmetry. In addition to $s$ exchange it is clear that the standard correlated two-pion 
exchange with pseudo-vector $\pi NN$ coupling remains (see Fig. 3c). It undergoes  the medium 
modifications 
of the $\pi\pi$ T-matrix discussed in section 2. This effect has been worked out in \cite{DKW93}. 
The overall change of the $NN$ potential depends very much on the relative 
weight of the two components, $s$ exchange and correlated $2\pi$ exchange,  
{\it i.e.}, on the sigma mass.
\begin{figure}
\begin{center}
\epsfig{file=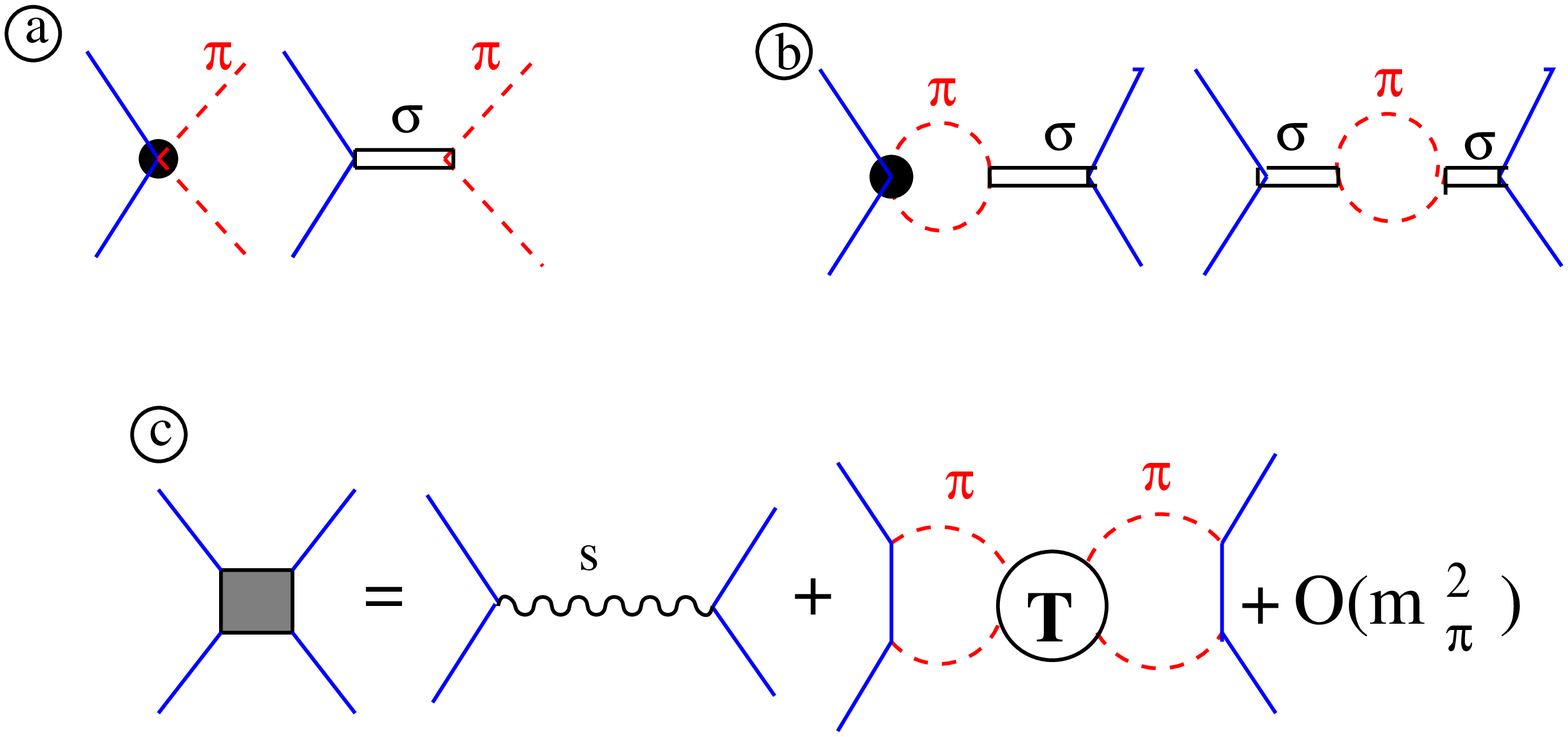,width=12.0cm,height=6.0cm,angle=0}
\end{center}
  \caption{ a) Compensating contributions to the $\pi N$ amplitude with pseudo-scalar coupling; 
  the blob represents the $N\bar N$ intermediate state. b) Corresponding compensation in the $NN$
  interaction leading to the suppression of the $2\pi$ dressing of the $\sigma$ propagator.
  c) Resulting $NN$ potential with undressed $\sigma$ exchange, {\it i.e.} $s$ exchange, and correlated
  two-pion exchange with in-medium modified $\pi\pi$ T-matrix (here the intermediate states are
  nucleon or $\Delta $ ones).}
\label{}      
\end{figure}  
 \section{Two components description of the $\sigma $ propagator and conclusion}
\begin{figure}
\begin{center}
\includegraphics[clip,height=3.0cm]{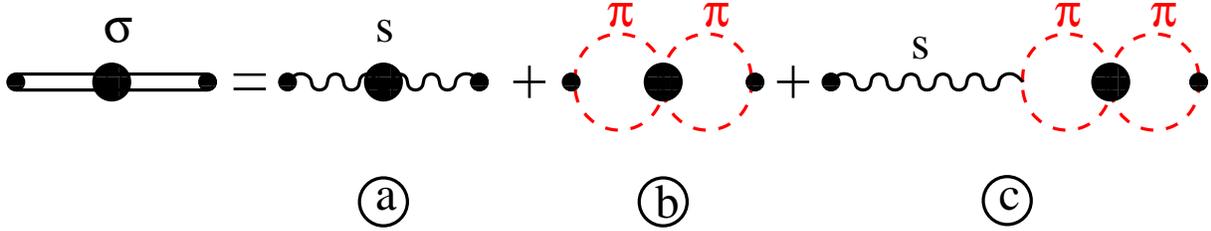}
\caption{\label{fig:sigma_s}Decomposition of the dressed $\sigma$ propagator into: a) the 
dressed $s$ propagator, b) the dressed two-pion propagator and c) the mixed term with a bare 
$s$ propagator and a dressed two-pion one.}
\end{center}
\end{figure}

In this last section we emphasize the separation of the $\sigma$ propagator 
with its strong  coupling to $2\pi $ states, into two components which
 are weakly coupled to each other. The decomposition writes
\begin{equation}
 D_{\sigma}(E)= D_s(E) +{3 \over 2 f^2_{\pi}}
\Big(1- 2{E^2~-m^2_{\pi}\over E^2~-m^2_{\sigma}} ~\Big) \tilde{G}(E) 
\end{equation}
where  $\tilde G$ is the fully dressed $2\pi$ propagator  
which obeys the equation~:
\begin{equation}
\label{GTILDE}
\tilde G=G\,+\,{1\over 2}\,\,G\,V\,\tilde G
\end{equation}
where $V$ has being given in Eq.(\ref{V}). 
The physical interpretation of the relation (\ref{GTILDE}) is simple and it is illustrated in Fig. \ref{fig:sigma_s}. 
The sigma propagator contains three parts~: the dressed s propagator with its derivative coupling 
to two-pion states (4 a), the fully dressed two-pion propagator (4 b), the third part is made of mixed terms 
with a bare $s$ propagator $D^0_s~=~(E^2-m^2_\sigma)^{-1}$ coupled to a fully dressed two-pion 
states (4 c). The coupling of the $s$ to two-pion states being $(E^2-m^2_\pi)$
 this leads to the third term (second term in the factor of $\tilde G$).
The two-pion propagator is the only piece which survives when $m_\sigma$ becomes infinite. To order
$m_\pi^2$, the medium effects that we have introduced in this work appear only in 
the two-pion component.
In the $NN$ interaction, which corresponds to the situation $E=0$ 
we have shown that this component, is eliminated and only the 
first one, which to order $m^2_\pi$ is not renormalized, is active.\\

In summary the unifying theme of the present work has been the question of the two-pion propagator in 
the nuclear medium. It is
modified by the dressing of the pionic lines by particle-hole bubbles. 
The $2\pi$ propagator appears in many problems. One of them, already explored, 
is related to the $2\pi$ production experiments on nuclei. It enters through its influence on
 the $\pi\pi$ T-matrix and its nuclear modification is responsible for the softening of the scalar strength
 observed in these data [12,25-27].\\
 In the present work we have focused on the case of zero four momentum for the two-pion propagator since this kinematical situation corresponds 
 to the problem  of the QCD scalar susceptibility. 
 Here the dressing of the pion lines
 by $p-h$ bubbles introduces in the nuclear susceptibility the response of the individual nucleons to a change of 
 the light quark mass. This response is dominated by the nucleon pion cloud. 
 As the cloud is polarized in the medium we have shown that this response
 undergoes a large renormalization in the sense of an enhancement of its magnitude, which can reach typically $\simeq$ 50$\%$ at $\rho=\rho_0$. 
 The contribution of the nucleon scalar susceptibility
 brings the nuclear one closer to the pseudoscalar one. Thus the nuclear pions have a double role in the restoration
 effects: first they participate in the decrease of the quark condensate, \textit{i.e.} also of the pseudoscalar
 susceptibility which follows the condensate. In addition through the softening of the scalar strength they increase the scalar susceptibility, 
 thus participate further in
 the convergence between the two susceptibilities, which then become nearly equal at the ordinary
 density, a remarkable convergence effect which signals chiral symmetry restoration.\\
 There is indeed a message about chiral symmetry
restoration contained in the $2\pi$ production experiments on nuclei. A softening of a strength, as is observed,
naturally translates into an increase (in magnitude) of the corresponding scalar
susceptibility, which is the inverse energy weighted sum rule.
 
 Another quantity to be influenced by the 2$\pi$ propagator is the propagator
 of the sigma meson, chiral partner of the pion, which is coupled to two-pion states. 
We have shown that, at zero four momentum, the change of the $2\pi$ propagator in the medium produces a 
major modification
 of the $\sigma$ propagator which can triple its magnitude.\\ 
 Our next step
 has been the investigation of the consequences for the $NN$ interaction in the nuclear medium.  
The scalar meson responsible for the nuclear attraction
 cannot be the previous sigma meson, but it has to be a chiral invariant scalar field. We had shown
 in a previous work that such an object can be found in the linear sigma model, being related to the 
fluctuation of the radius of the chiral circle. Contrary to the $\sigma$ this invariant field is weakly coupled to the pion. 
Therefore it does not undergo the large medium modification of the $\sigma$, 
insuring the stability of the $NN$ interaction in the nuclear medium.



\begin{thebibliography}{99}
\bibitem{B00} F. Bonnuti {\it et al.}, Nucl. Phys. {\bf A677} (2000) 213; 
Phys. Rev. Lett. {\bf  77} (1996) 603; Phys. Rev. {\bf C60} (1999) 018201. 
\bibitem{ST00} A. Starostin {\it et al.}, Phys. Rev. Lett. {\bf 85} (2000) 
5539.
\bibitem{M02}J.G. Messchendorp {\it et al.}, Phys. Rev. Lett. {\bf 89} 
(2002) 222302.
\bibitem{Krusche:2004uw}
  B.~Krusche {\it et al.},
  Eur.\ Phys.\ J.\  {\bf A22} (2004) 277.
\bibitem{SNC88} P. Schuck, W. Norenberg and G. Chanfray, Z. Phys. {\bf A330} (1988)
119. 
\bibitem{CASN91} G. Chanfray, Z. Aouissat, P. Schuck and W. Norenberg, 
Phys. Lett. {\bf  B256} (1991) 325.
\bibitem{HKS99} T. Hatsuda, T. Kunihiro and H. Shimizu, Phys. Rev. Lett. 
{\bf 82} (1999) 2840.
\bibitem{JHK01} D. Jido, T. Hatsuda and T. Kunihiro, Phys. Rev. {\bf D63} (2001) 011901.
\bibitem{MABM04} P. Muhlich, L. Alvarez-Ruso, O. Buss and U. Mosel, 
                 Phys. Lett. {\bf B595} (2004) 216.
\bibitem{CEG02} G. Chanfray, M. Ericson and P.A.M. Guichon, Phys. Rev 
{\bf C63} (2001) 055202.
\bibitem{ASW97} Z. Aouissat, P. Schuck and J. Wambach, Nucl. Phys. {\bf A618} (1997) 402.
\bibitem{ROV02} L. Roca, E. Oset and M.J. Vicente Vacas, Phys. Lett. {\bf B541} 
(2002) 77.
\bibitem{ChEr93} G. Chanfray and M. Ericson, Nucl. Phys. {\bf A556} (1993) 427.
\bibitem{CEG03} G. Chanfray, M. Ericson, and P.A.M. Guichon, Phys. Rev 
{\bf C68} (2003)  035209.
\bibitem{Ose87}         E. Oset and L.L Salcedo, 
                        Nucl. Phys. {\bf A468} (1987)  631.
\bibitem{Chanfray:1999nn}
  G.~Chanfray and D.~Davesne,
  Nucl.\ Phys.\  {\bf A646} (1999) 125.
\bibitem{CE03} G. Chanfray and M. Ericson, Eur.\ Phys.\ J.\ {\bf A16} (2003) 291.
\bibitem{AcHu} G. Chanfray and M. Ericson, Acta Phys. Hung. A 19/1 (2004), Heavy Ion Physics.
\bibitem{CE04} G. Chanfray and M. Ericson, nucl-th/0402018, to be published on Eur.\ Phys.\ J.\ A.
\bibitem{ARCSW95} Z. Aouissat, R. Rapp, G. Chanfray, P. Schuck, J. Wambach, Nucl. Phys. 
{\bf A581} (1995) 471.
%
\bibitem{B96} M Birse, Phys. Rev. {\bf C53} (1996) 2048; Acta Phys. Pol. {\bf B29} (1998)
2357.
\bibitem{SW86} B.D. Serot and J.D. Walecka, Adv. Nucl. Phys. {\bf 16} 
(1986) 1; Int. J. Mod. Phys. {\bf E16} (1997) 515.
\bibitem{B94} M. Birse, Phys. Rev. {\bf C49} (1994) 2212.
\bibitem{DKW93} J.W.Durso, H.C.Kim and J. Wambach, Phys. Lett. {\bf B298}
 (1993) 267.
\bibitem{Rapp:1998fx}
R.~Rapp, J.W.~Durso, Z.~Aouissat, G.~Chanfray, O.~Krehl, P.~Schuck, J.~Speth and J.~Wambach,
Phys.\ Rev.\  {\bf C59} (1999) 1237. 
\bibitem{Schuck:1998nw}
P.~Schuck, Z.~Aouissat, F.~Bonutti, G.~Chanfray, E.~Fragiacomo, N.~Grion and J.~Wambach,
arXiv:nucl-th/9806069.
\bibitem{VO99} M.J. Vicente Vacas and E. Oset, Phys. Rev. {\bf C60} (1999) 064621.

\end{thebibliography}
\end{document}